\documentclass[fleqn,usenatbib]{mnras}

\usepackage{mathptmx}
\usepackage[T1]{fontenc}
\usepackage{ae,aecompl}

\usepackage{graphicx}	% Including figure files
\usepackage{amssymb}	% Extra maths symbols

\def\asec{\ifmmode ^{\prime\prime}\else$^{\prime\prime}$\fi}

\def\it{\sl}
\def\degs{\ifmmode ^{\circ}\else$^{\circ}$\fi}
\def\amin{\ifmmode ^{\prime}\else$^{\prime}$\fi}
\def\asec{\ifmmode ^{\prime\prime}\else$^{\prime\prime}$\fi}
\def\farcs{\hbox{$.\!\!^{\prime\prime}$}}  % Fractions of arcseconds
\def\psr{PSR~J1357$-$6429}
\def\degs{\ifmmode ^{\circ}\else$^{\circ}$\fi}
\def\amin{\ifmmode ^{\prime}\else$^{\prime}$\fi}

\def\eqalign#1{\null\,\vcenter{\openup1\jot \m@th
   \ialign{\strut\hfil$\displaystyle{##}$&$\displaystyle{{}##}$\hfil
   \crcr#1\crcr}}\,}
\title[Near-IR observations of PSR J1357$-$6429]{Near-IR observations of PSR J1357$-$6429}

\author[D.~Zyuzin et al.]{D.~Zyuzin,$^{1}$\thanks{E-mail: da.zyuzin@gmail.com (DZ)}
S.~Zharikov,$^{2}$
Yu.~Shibanov,$^{1,3}$
A.~Danilenko,$^1$
R.~E.~Mennickent$^{4}$ and
\newauthor
A.~Kirichenko$^{1}$ 
\\
$^{1}$Ioffe Institute, 26 Politekhnicheskaya st., St. Petersburg 194021, Russia\\
$^{2}$Observatorio Astron\'{o}mico Nacional SPM, Instituto de Astronom\'{i}a, Universidad Nacional Aut\'{o}noma de Mexico, Ensenada, BC, Mexico\\
$^{3}$Peter the Great St. Petersburg Polytechnic University, 29 Politekhnicheskaya st., St. Petersburg 195251, Russia\\
$^{4}$Department of Astronomy, Universidad de Concepcion, Casilla 160-C, Concepcion, Chile\\
}

\date{Accepted XXX. Received YYY; in original form ZZZ}

\pubyear{2015}

\begin{document}
\label{firstpage}
\pagerange{\pageref{firstpage}--\pageref{lastpage}}
\maketitle

\begin{abstract}
PSR J1357$-$6429 is a young radio pulsar that was detected in
X-rays and $\gamma$-rays. We present the high spatial resolution near-infrared imaging of the pulsar
field in $J$, $H$ and $K_s$ bands obtained with the VLT/NaCo using the Adaptive Optic system.
We found a faint source at the  most precise pulsar
radio position which we propose as the pulsar near-infrared counterpart
candidate. It is confidently detected in the $J$ and $K_s$ bands, with $J$ =
23.51$\pm$0.24 and $K_s$ = 21.82$\pm$0.25.
There is a hint of the source in the $H$ band with an upper limit $H$ $>$ 22.8.
The dereddened source fluxes are compatible with the extrapolation of the
pulsar X-ray spectrum towards the near-infrared.
If the candidate is the true counterpart, by this property PSR J1357$-$6429 would be 
similar to the nearby middle-age pulsar PSR B0656+14. In this case, both
pulsars demonstrate an unusually high near-infrared efficiency relative to the
X-ray efficiency as compared to other pulsars detected in both ranges.
\end{abstract}

\begin{keywords}
%keyword1 -- keyword2 -- keyword3
Pulsars: individual: PSR J1357$-$6429
\end{keywords}

%%%%%%%%%%%%%%%%%%%%%%%%%%%%%%%%%%%%%%%%%%%%%%%%%%

%%%%%%%%%%%%%%%%% BODY OF PAPER %%%%%%%%%%%%%%%%%%

\section{Introduction}
\label{Sect:intro}
A young radio pulsar J1357$-$6429 (hereafter J1357) 
has a characteristic age $\tau$ = 7.3 kyr, a period $P$ = 166 ms,
a spin-down luminosity $\dot{E}$ = 3.1 $\times$ 10$^{36}$ ergs~s$^{-1}$  
and a distance of $\approx$ 2.5 kpc derived from the dispersion measure 
\citep{Camilo04}. \textit{XMM-Newton} and \textit{Chandra}  
X-ray observations have revealed the pulsar X-ray counterpart and 
the pulsar wind nebula (PWN) containing   
a tail-like structure extended to several tens of arcseconds 
from the pulsar and a fainter plerion extended up to a few tens of arcminutes 
\citep{Esposito07,Zavlin07,Chang2012,Lemoine-Goumard11}.
The plerion was detected in the TeV range with the High Energy Stereoscopic System (H.E.S.S.) and 
was also found in archival radio data \citep{Abramowski11}.
Based on the 7\amin\ offset between the pulsar and the plerion centre,  \citet{Abramowski11} estimated
a high pulsar transverse velocity of 650$d_{2.5}\tau_{7.3}^{-1}$ km s$^{-1}$, where 
$\tau_{7.3}$ is the pulsar age in units of 7.3 kyr and $d_{2.5}$ is the distance 
to the pulsar in units of 2.5 kpc. 
However, \citet{Abramowski11} also noted that the plerion can be displaced by the 
reverse supernova remnant shock that had propagated
through the non-homogeneous supernova remnant interior \citep{Blondin2001}. 
In that case the pulsar velocity may be much smaller.

Pulsations of J1357 with the pulsar period were 
discovered in X-rays with \textit{XMM-Newton} 
and in the GeV range with \textit{Fermi} \citep{Lemoine-Goumard11,Chang2012}. 
The pulsar X-ray spectrum shows 
a thermal component originating from the surface of the neutron star (NS) 
and a non-thermal power-law component describing the pulsar 
magnetosphere emission \citep{Lemoine-Goumard11,Chang2012, Danila12}.  

The J1357 field was observed in the optical $VRI$ bands with  
the ESO Very Large Telescope (VLT) in 2009 \citep{Mignani11, Danila12}. \citet{Danila12}  
detected  a point source in all these bands 
within 1$\sigma$ uncertainties of the pulsar X-ray position
and suggested it as a pulsar optical counterpart candidate.
The 5$\sigma$ offset of the source from the pulsar radio interferometric position, which was 
obtained $\approx$9 years earlier \citep{Camilo04}, 
implied a very high pulsar transverse velocity of $\approx$ 2000$d_{2.5}$ km s$^{-1}$ \citep{Danila12}. 
Similar velocity was obtained by \citet{Mignani11} based on the comparison of the pulsar radio and X-ray  positions.
However, \citet{kirichenko2015arXiv} showed that \citet{Camilo04} underestimated the 
uncertainties of the  pulsar radio position obtained with the Australia Telescope Compact Array (ATCA) in 2000.
\citet{kirichenko2015arXiv} also presented results of the new ATCA observations  performed in 2013.
They found no shift of the pulsar at the 13 yr timebase between the two radio observations and put a new 
upper limit on the pulsar transverse velocity 
of $\lesssim 1200d_{2.5}$ km~yr$^{-1}$. 
Given a higher accuracy of the new ATCA pulsar position as compared to  
the X-ray position, 
they also discarded the pulsar optical counterpart proposed by \citet{Danila12}. 

Here we report first deep near-infrared (IR) observations 
of the J1357 field with the VLT using the Adaptive Optics (AO) system.
High spatial resolution provided by the AO system allowed us to find a new pulsar near-IR counterpart candidate 
which is unresolved 
from a bright star located within 1\asec\ of the pulsar radio position in the previous seeing-limited VLT optical images. 
The observations and data reduction are described  
in Sect.~\ref{sec:obs}, our results are presented in Sect.~\ref{sec3}  
and discussed  in  Sect.~\ref{sec4}.

\section{The VLT data}
\label{sec:obs}

\subsection{Observations and data reduction}
%\label{sec:obs}
The pulsar field was observed in the $J$, $H$, and $K_s$ bands
with the  Nasmyth Adaptive Optics System (NAOS) and Near-Infrared Imager and Spectrograph (CONICA)
(in short NaCo) at the VLT/UT4 unit in the period of 2012--2013. 
The S27 instrument mode was used with an image scale of 0\farcs027/$\mathrm{pixel}$ 
and a field of view  (FOV) of $\sim$27\asec$\times$27\asec. A bright 
natural star of $V\approx$13.8 located  $\approx$11\farcs2 from the 
target was used for  AO wavefront corrections with 
the visual NAOS Dichroic Wavefront sensor. 
Several sets of 200 s dithered exposures were obtained in each band. 
After a preliminary inspection, we 
selected only sets where the observing conditions were rather 
stable and photometric, with  seeing values varying from 0\farcs7 to 0\farcs9 
 (Table~\ref{t:log}). 

%%%%%%%%%%%%%%%%%%%%%%%%%%%% Table 1 %%%%%%%%%%%%%
\begin{table} %[]
\caption{Observations of \psr\ with the VLT/NaCo under the Program 089.D-0956A.}
\begin{center}
\begin{tabular}{llcll}
\hline
   Date           &  Band            & Exposure       &  Mean          & Seeing          \\
                  &                  & $\times$ exp. number                &  airmass       & range           \\
                  &                  & [s]            &                & [arcsec]      \\
\hline                
   2013-04-17     &  $J$           & 200$\times$10            & 1.31          &  0.70--0.80       \\
   2013-03-15     &  $H$           & 200$\times$10            & 1.32           &  0.83--0.89       \\                              
   2013-04-29     &  $K_s$         & 200$\times$10            & 1.31           &  0.69--0.88       \\
\hline
\end{tabular}
\end{center}
\label{t:log}
\end{table}

We performed standard near-IR data reduction, including  bias subtraction, 
flat-fielding, sky-subtraction and cosmic-ray
removal using the {\tt IRAF} \textit{xdimsum} package.  
Resulting full widths at half maximum (FWHM) of a point source 
near the target position on the combined $J$, $H$ and $K_s$ images were  
$\approx$ 0\farcs10,  0\farcs13 and 0\farcs08 at 
mean airmass of $\approx$1.3 for each band.    
Total integration time for each of the combined images was 2 ks. 

%%%%%%%%%%%%%%%%%%%%%%%%%%%%%%%%%%%%%%%%%%%%%
\subsection{Astrometry} 
%%%%%%%%%%%%%%%%%%%%%%%%%%%%%%%%%%%%%%%%%%%%%

As a reference frame for astrometric referencing, we used the VLT  
$I$-band image with a much  
wider FOV and an absolute astrometric referencing 
uncertainty of $\la 0\farcs2$   obtained by \citet{Danila12}.  
We selected fifteen common isolated unsaturated   
stars as secondary astrometric standards in this frame 
and in the best-quality near-IR  resulting image 
obtained in the $K_s$ band. 
Their absolute WCS coordinates were found using the optical frame. 
We then derived their pixel coordinates in the $K_s$ image 
making use of the {\tt IRAF} task {\it imcenter} 
with an accuracy of $\la$ 0.025 of the image pixel. 
The {\tt IRAF} tasks {\sl ccmap/cctran} were applied 
to the astrometric transformation of the $K_s$ image.  
Formal {\sl rms} uncertainties of the  astrometric
fit  were $\Delta$RA $\la$ 0\farcs05 and $\Delta$Dec. $\la$ 0\farcs02.
The resulting $J$ and $H$ images were  aligned to the $K_s$  
frame with an accuracy of  $\la$ 0\farcs02.
Combining all uncertainties and accounting for the optical 
astrometric referencing accuracy,
a conservative estimate of our near-IR astrometric 
referencing uncertainty  is  $\la$ 0\farcs21 for 
both coordinates in all three bands.

%%%%%%%%%%%%%%%%%%%%%%%%%%%%%%%%
\subsection{Photometric calibration}                
%%%%%%%%%%%%%%%%%%%%%%%%%%%%%%%%%%%%%%%%%%%%%%%%%%%%%%
\label{photcal}
For the photometric calibration, standards 9157, 9144 and 9150 from \citet{Persson98} 
were observed during the same nights as our target. 
We fixed the atmospheric extinction coefficients
at their mean values adopted from the VLT
homepage\footnote{http://www.eso.org/sci/facilities/paranal/ \\ decommissioned/isaac/tools/imaging\_standards.html}:
k$_J$ = 0.11, k$_H$ = 0.06 and k$_{K_{s}}$ = 0.07. 
As a result,  we obtained the following magnitude zero-points for
the summed images: 
$J^{ZP}$ = 24.03 $\pm$ 0.02, $H^{ZP}$ = 23.87 $\pm$ 0.01 
and $K_{s}^{ZP}$ = 22.99 $\pm$ 0.01.  
The $3\sigma$ detection limits for a point-like object 
within several arcseconds of the target position 
are $J$ $\approx$ 24.7, $H$~$\approx$~23.2 and $K_s$ $\approx$ 22.7.

%%%%%%%%%%%%%%%%%%%%%%%%%%%%%%%%%%%%%%%%%%%%
\section{Results}
\label{sec3}
\subsection{Searching for a pulsar near-IR counterpart}
%%%%%%%%%%%%%%%%%%%%%%% Fig 1  %%%%%%%%%%%%%%%%%%%%%%%%%%%%%%%%%%%%%%%%%%%%%%%%%%%%%%%%%%
\begin{figure} %[htp]
\setlength{\unitlength}{1mm}
\begin{center}
\includegraphics[width=70mm, clip=]{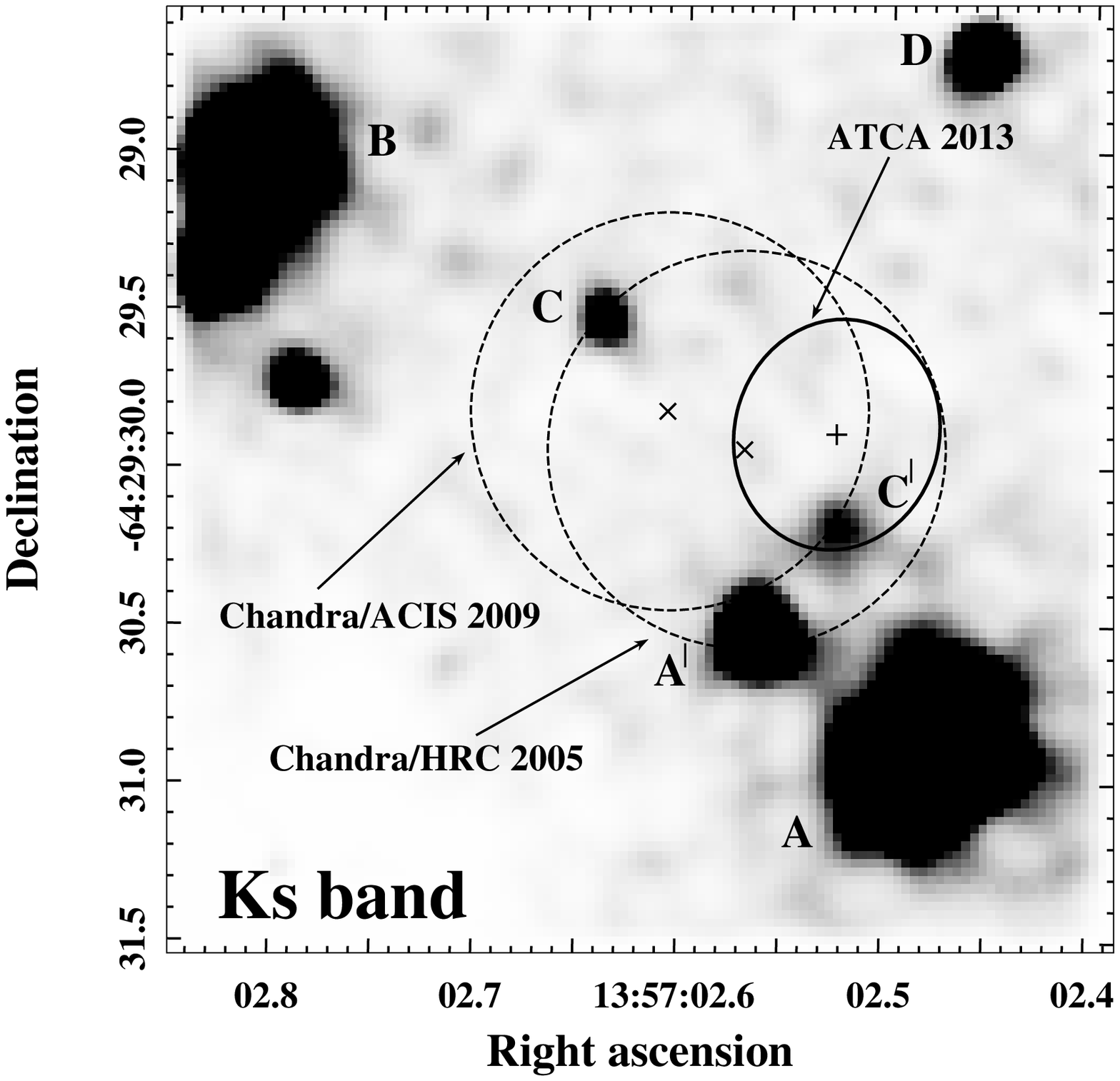}
\includegraphics[width=70mm, clip=]{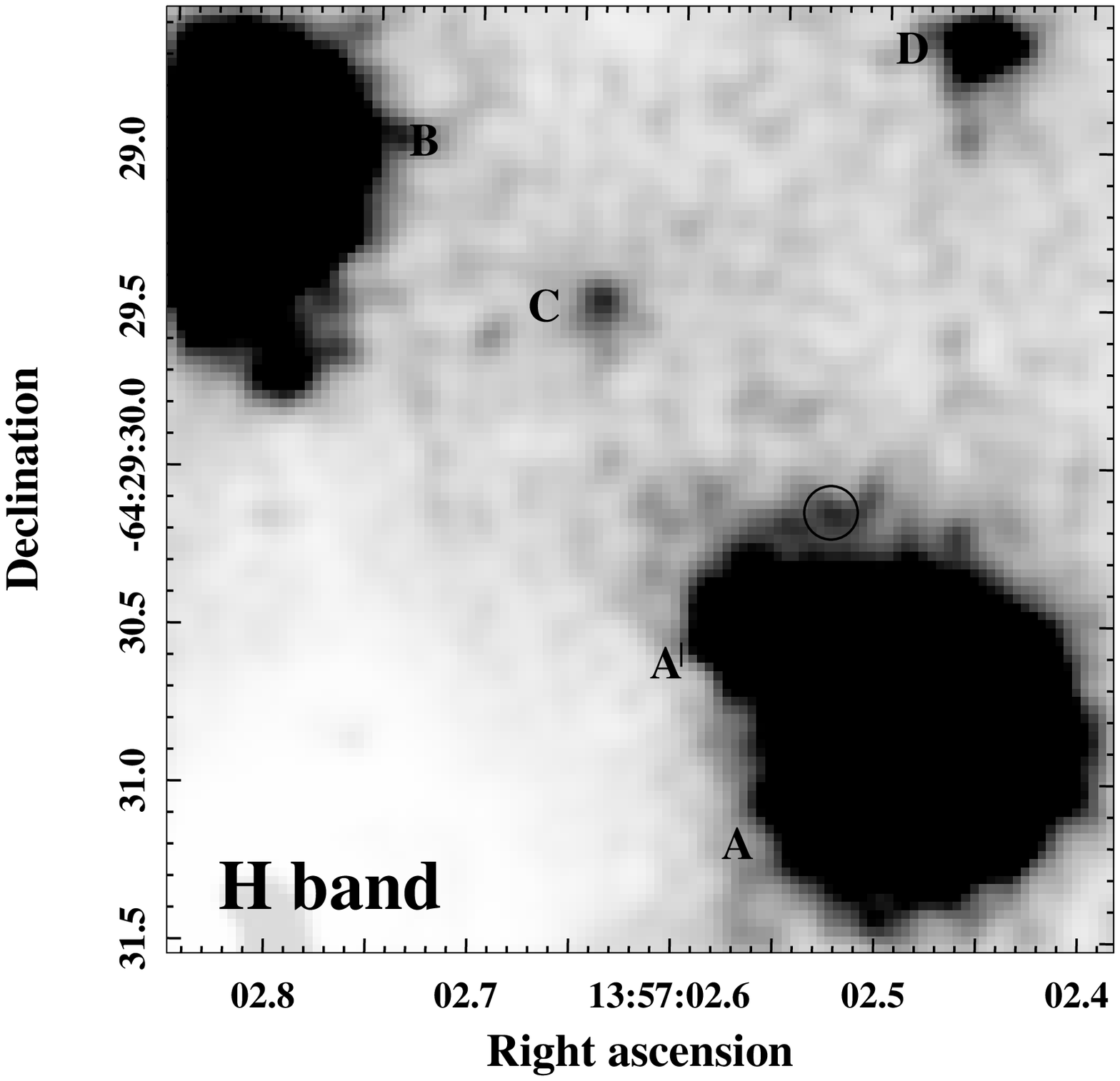}
\includegraphics[width=70mm, clip=]{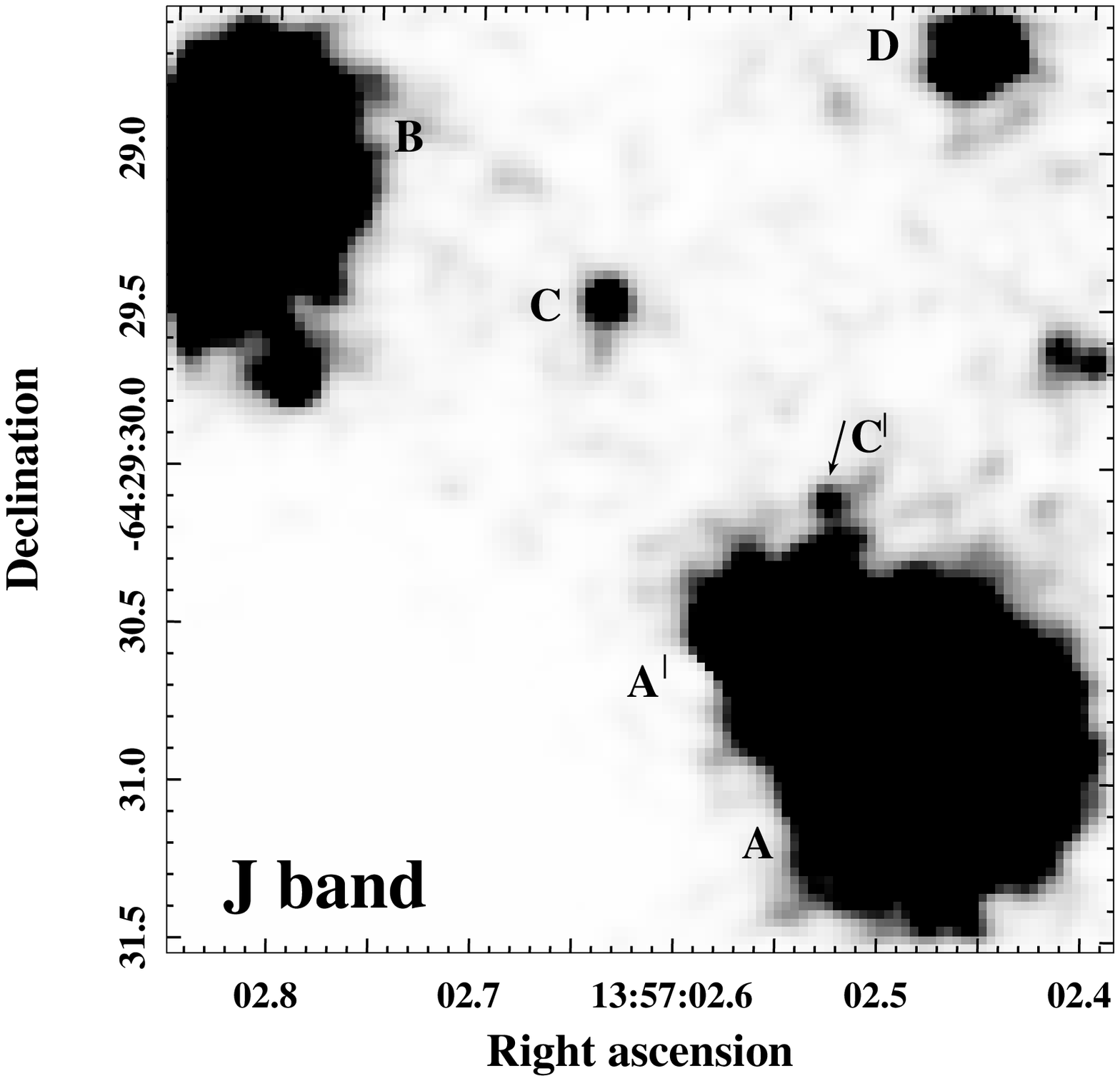}
\end{center}
 \caption{ $3\asec\times3\asec$ image fragments of the J1357 field 
 obtained with the VLT/NaCo  in $K_s$ (\textit{top}), 
 $H$ (\textit{middle}),  and $J$ (\textit{bottom}) 
 bands.   The $X$-points with dashed-line circles  and  the $+$ with the thick-line ellipse 
 in the \textit{top} panel  show the pulsar positions and their 1$\sigma$ uncertainties 
 from the two  \textit{Chandra} X-ray 
 and  new ATCA radio observations, respectively.
 $A$, $A'$, $B$, $C$, and $D$ label 
 objects as in \citet{Danila12}.
 $C'$ labels the pulsar new counterpart candidate. 
 It is not detected in the $H$ band owing to a worse FWHM, although  
 some flux excess is seen on its position marked by the circle.}
\label{fig:1}
\end{figure}
%%%%%%%%%%%%%%%%%%%%%%%%%%%%%%%%%%%%% end Fig-1 %%%%%%%%%%%%%%%%%%%%%%%%%%%%%%%%%%%%%%%%

The fragments of the resulting $JHK_s$ images containing the pulsar are presented in Fig.~\ref{fig:1}. 
All sources detected and studied earlier by \citet{Danila12} in the optical bands 
are marked following their alphabetic nomenclature, including   
the object $C$  which was proposed as a possible pulsar optical counterpart. 
All these objects are resolved  better than in the optical.  
The X-ray positions of the pulsar obtained from \textit{Chandra}/HRC and \textit{Chandra}/ACIS 
observations performed in 2005 and  2009, respectively, 
and the new pulsar radio position obtained by \citet{kirichenko2015arXiv}
are shown in the $K_s$ image together with their 1$\sigma$ error ellipses.   
The ellipses account for
the uncertainties of the X-ray/radio position measurements and the near-IR astrometry.
As seen, the  most precise coordinates of the pulsar are based on the radio observations:  
RA = 13:57:02.525(14) and Dec. = $-$64:29:29.89(15)\footnote{Herein, 
the numbers in brackets are  1$\sigma$ uncertainties referring  
to the last significant digits quoted.} \citep{kirichenko2015arXiv}. 

At this position we detected a new potential pulsar counterpart, 
a faint object marked by $C'$ in Fig.~\ref{fig:1}. 
It also overlaps  with the  1$\sigma$  error ellipses of the  pulsar X-ray positions.  
The source coordinates are RA=13:57:02.52 and Dec.=$-$64:29:30.15.    
Its 1$\sigma$ uncertainty for both coordinates is $0\farcs22$. 
The object is  detected  in  the $J$ and $K_s$ bands 
at $\approx4.5\sigma$ significance. 
There is also a hint of the source  in the $H$ band, 
but a factor of 1.5 as worse FWHM in this band (Sect.~\ref{sec:obs}) 
does not allow for a confident detection. 
In the   VLT optical images obtained with the seeing-limited  0\farcs5 spatial resolution,    
the object $C'$ is completely hidden in the spatial profile wings of relatively 
bright nearby stars $A$ and $A'$ \citep[see, e.g., fig. 9 from][]{kirichenko2015arXiv}. 

As in the optical \citep{Danila12},  we did not find any near-IR counterpart for the extended X-ray emission 
of the J1357 PWN.  

\subsection{Photometry}
\label{sec:phot}
We performed  aperture photometry of all point-like sources labelled in Fig.~\ref{fig:1}. 
The respective aperture corrections were measured and applied based 
on the photometry of bright unsaturated field stars located near the pulsar position. 
We also derived a 3$\sigma$ upper limit on the object $C'$ brightness in the $H$ band.
It is not as deep as the upper limit claimed in Sect.~\ref{photcal} due to 
the contribution  of $A$ and $A'$ star wings to  backgrounds at the $C'$ position.    
The results are collected in Table~\ref{t:mag}.  
The errors include  the statistical  measurement errors, 
calibration zero-points and aperture correction  uncertainties.
For completeness we also estimated upper limits on the object $C'$ brightness in the VLT 
$VRI$ images obtained by \citet{Danila12}: 
$V\ga24.6$, $R\ga24.2$ and $I\ga23.2$.

% L7-L8
% M(Ks)=11.9   21.33
%%%%%%%%%%%%%%%%%%%%%%%%%%%% Table 3 %%%%%%%%%%%%%
\begin{table} %[b]
\caption{Measured $JHK_s$ magnitudes of the point-like objects
detected in the J1357 vicinity and labelled in Fig.~\ref{fig:1}.}
\begin{center}
\begin{tabular}{llll}
\hline\hline
 Source               &  $J$                &  $H$       &  $K_s$                    \\
%                   &     mag,            &  mag,           &   mag,          \\
%                  &     $F_J$            &  $F_H$                &   $F_{K_{s}}$                            \\
   \hline \hline               
%   $C$     &    22.92(12)           & 22.06(15)           &   21.33(15)                                  \\
%   $C'$    &    23.51(24)           &   $>$22.8          &   21.82(25)                    \\
%   \hline 
   $A$       &   18.84(1)           &    18.15(1)        &    17.78(1)                    \\
   $A'$      &   20.89(3)           &    20.02(4)        &    19.68(3)                    \\
   $B$       &   19.43(1)           &    18.67(1)        &    18.32(1)                    \\
   $C$     &    22.92(12)           & 22.06(15)           &   21.33(15)                                  \\
   $C'$    &    23.51(24)           &   $\ga$22.8          &   21.82(25)                    \\
  $D$       &   21.89(5)           &    20.88(5)        &    20.38(6)                    \\
\hline
\hline
%$A_V=1.84$  & $A_J=0.52$  &  $A_H = 0.33$ & $A_{Ks} = 0.22$   \\
%$C'$  & 22.99(24) &  $>$22.47  &  21.60(25)   \\
%\hline
%$A_V=0.84$  & $A_J=0.24$  &  $A_H = 0.15$ &  $A_{Ks} = 0.10$   \\
%$C'$  & 23.27(24) &  $>$22.65 &  21.72(25)    \\
\end{tabular}
\end{center}
\label{t:mag}
\end{table}

%%%%%%%%%%%%%%%%%%%%%%%%%%%%%%%%%
\section{Discussion} \label{sec4}  
%%%%%%%%%%%%%%%%%%%%%%%%%%%%%%%%%
        
Considering  potential optical/IR pulsar counterparts,  
we adopt the pulsar new radio interferometric 
position derived by \citet{kirichenko2015arXiv} as the most precise one. 
The radio and near-IR observations were performed 
almost simultaneously.   
The only object within the 1$\sigma$ error ellipse of this position is the source $C'$ (Fig.~\ref{fig:1}).  
Based on the X-ray data \citep[see][]{Danila12} a true counterpart  can hardly  
be brighter than $K_s\approx 20.5$.    
In the NaCo FOV the surface  number density of observed point-like objects 
with $K{_s}\ga20.5$ down to the  limiting magnitude of 22.7 
is $\approx$0.16 object arcsec$^{-2}$. 
The confusion probability to find an unrelated  source of such brightness 
within the radio  error ellipse with the area of $\approx 0.5$ arcsec$^{-2}$ 
is $\approx$0.08. Below we   
discuss the source $C'$  as the counterpart candidate.  

\begin{figure} 
\begin{center}
\includegraphics[scale=0.35, clip]{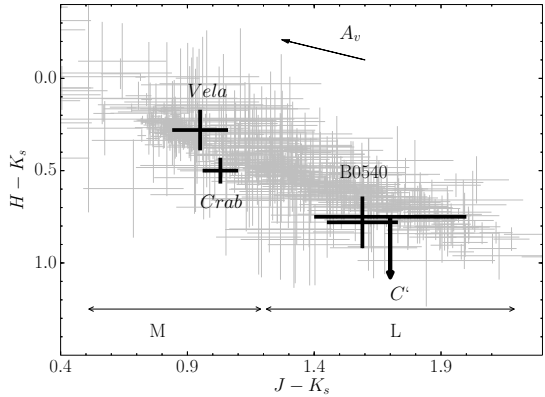}
\end{center}
\caption{Colour--colour diagram of the M and L dwarfs and the observed colours of the object $C'$. % (red). 
The ranges of the L and M %, and T 
dwarf colours are shown by double headed arrows with respective labels.     
A dereddening vector marked by ``Av''
shows the direction and length of the shift for the object $C'$ 
on the diagram for A$_{\rm V}$ = 2. For comparison, the colours of the Vela, Crab and B0540$-$69 pulsars
are shown by error-bars.}
\label{fig:col-col}
\end{figure}
%%%%%%%%%%%%%%%%%%%%%%%%%%%%%%%%%%%%%%%%%%%%%%%%%%%%%%%%%%%%%%%%%%%%%%%%

%%%%%%%%%%%%%%%%%%%%%%%%%%%% Fig 2 %%%%%%%%
\begin{figure} %[t]
\begin{center}
\includegraphics[scale=0.35, clip]{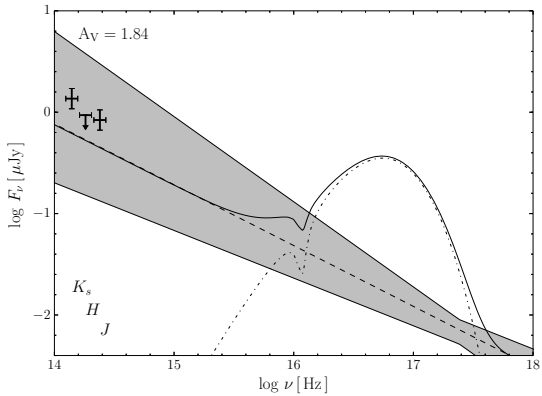}
\end{center}
 \caption {Tentative multi-wavelength unabsorbed spectrum of J1357. The X-ray part is described by 
 the magnetized NS hydrogen  atmosphere and
 power law models and is  
 taken from \citet{Danila12}. The solid, dashed and dash-dotted lines are the total X-ray spectrum and 
 contributions of the atmospheric and power law components, respectively. The grey region shows uncertainties 
 of the power law component. The X-ray spectrum is extrapolated to the near-IR.
 The dereddened $JK_s$ fluxes and the $H$-band upper limit for the object $C'$ 
 are shown by bold error bars.}
\label{fig:mw1}
\end{figure}

Stellar colour-magnitude and colour-colour diagrams are frequently  used to distinguish pulsars 
from field stars by their peculiar colours \citep[e.g.,][]{0656Durant, Mignani11, Danila12}. 
Using the 2MASS Data Mining and the M, L and T Dwarf Archives 
from \citet{2MASS-dwarfs}, we created the near-IR colour-colour diagram shown in Fig.~\ref{fig:col-col}.
The archival data are related to nearby dwarfs located within about 100 pc  where reddening 
is negligible. Therefore, the star colours are consistent with their intrinsic colours. 
The $J-K_s$ colour of the object $C'$ doesn't appear peculiar and 
would be compatible with  L-type dwarf colours (Fig.~\ref{fig:col-col}). 
However, as seen from Fig.~\ref{fig:col-col}, the colours for other pulsars detected in the near-IR, e.g., 
the Crab, Vela and PSR B0540$-$69, 
are also hardly distinguishable from  those of the M-L-dwarf sequence.
Therefore, the near-IR colour analysis  is not informative as it does not allow to confirm or disclaim 
the candidate as the pulsar counterpart.    

Another possibility to establish the $C'$ nature  is to compare its IR fluxes with the pulsar X-ray spectral data  
and to see if the compiled IR-X-ray spectral energy distribution (SED) is compatible with those for other 
pulsars observed in both ranges.  The tentative J1357 SED is presented in Fig.~\ref{fig:mw1}.   
The IR fluxes of  $C'$ are dereddened using the interstellar extinction $A_V\approx1.84$ towards the pulsar estimated 
by \citet{Danila12}. 
The unabsorbed X-ray spectrum of J1357 is reproduced using the best-fit spectral parameters from the same paper.
The pulsar X-ray radiation is equally well described by the magnetized NS  
atmosphere and blackbody spectral models for the thermal emission 
dominating at low energies and by the power law responsible for the high energy spectral tail.  
As an example,  in Fig.~\ref{fig:mw1} we show the result for the atmosphere model. 

As seen from Fig.~\ref{fig:mw1}, the  near-IR fluxes 
of the counterpart candidate are  consistent within 1$\sigma$ uncertainties 
with the long wavelength extrapolation of the pulsar  
X-ray spectrum. Replacing the atmosphere model with the blackbody one doesn't change this result.
Similar situation is observed for the middle-aged pulsar B0656$+$14 \citep{Shibanov06, 0656Durant}.   
For other pulsars detected in the near-IR, 
the IR fluxes are by a factor of 10--100 lower than  the X-ray extrapolation 
\citep[e.g.,][]{Mignani12-0540nIR, NustarGeminga}. Therefore, if $C'$ is the real counterpart of
J1357, by its multi-wavelength spectral properties this pulsar can be similar  to PSR B0656$+$14, 
which demonstrates an unusually high brightness in the near-IR.   
%%%%%%%%%%%%%%%%%%%%%%%%%%%%% Fig 2 %%%%%%%%
\begin{figure} 
 \begin{center}
  \includegraphics[scale=0.3, clip]{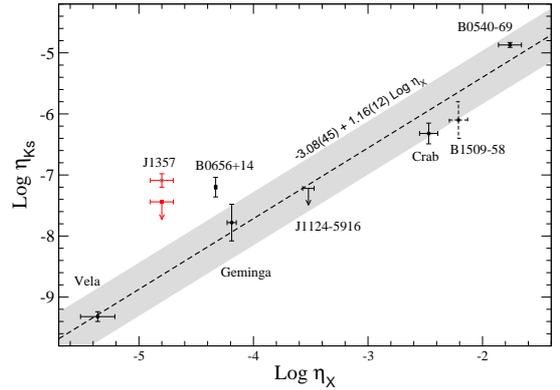}  
 \end{center}
 \caption{$K_s$-band efficiencies vs. X-ray efficiencies for pulsars observed in both ranges. The data are presented in Table~\ref{t:IR}. 
 The dashed line and grey region show the best fit and its 1$\sigma$ uncertainty to the data for four firmly detected 
 near-IR pulsars (marked by black circles) excluding B0656+14. J1357 is represented by the  near-IR counterpart candidate $C'$ 
 and by the detection limit of our observations. The near-IR counterpart candidate of PSR B1509-58 
 and the near-IR upper limit for PSR J1124-5916 are shown for comparison.}
 \label{fig:eff-eff}
\end{figure}
%%%%%%%%%%%%%%%%%%%%%%%%%%%%%%%%%%%%%%%%%%%%%%%%%%%%%%%%%%%%%%%%%%%%%%%%

\begin{table*}
\caption{Near-IR and X-ray parameters of pulsars detected in both ranges. 
The observed $K_s$ magnitudes and $A_V$ values for all pulsars except of J1357 and Vela are taken 
from \citet{zhar13}. The observed $K_s$ magnitude for Vela is taken from \citet{zyuzin13}.
X-ray data and other parameters of pulsars are adopted from \citet{kar-pav08}. 
IR data for J1357 are represented by the counterpart candidate $C'$ 
and by the detection limit of our observations. 
The near-IR flux upper limit for PSR J1124$-$5916 \citep{zhar13} and the data for the near-IR 
counterpart candidate of PSR B1509$-$58 \citep{Kaplan06-B1509-nIR} are included for generality.}
\begin{tabular}{llllllllll}
\hline\hline
Pulsar name	      & $\log \tau$ & $\log \dot{E}$ &  distance &  $A_V$  & K$_s$   & $\log L_{K_s}$   & $\log \eta_{K_s}$  & $\log L_{x}^{**}$   & $\log \eta_{x}$  \\ \hline \hline
	      & yr & erg s$^{-1}$ &  kpc & mag  & mag  &  erg s$^{-1}$  &     & erg s$^{-1}$   &   \\ %& Ref.
\hline
Crab	     &   3.1     &	38.65	  &  1.73(28)  & 1.62  &	 13.77(5)	 &  32.33(17)  &  $-$6.32(17) &36.19(1)  &  $-$2.47(1)  \\ % &	(1)	\\
 PSR 0540$-$69  &  3.2     &  38.17   &   50  &   0.62  &  18.55(10)  & 33.30(4)  & $-$4.87(4) & 36.41(10)  &  $-$1.76(10)  \\ %& (2)     \\
 PSR B1509$-$58  &   3.2     &  37.25     &  5.2(1.4)  & 4.8 & 19.4(1)& 31.18(30) &  $-$6.1(3) & 35.04(8)  &  $-$2.21(8)  \\ %&(3) \\ 
    
 PSR 1124$-$5916 &  3.5     &  37.08     &   $\sim6.0$  & 1.98(12)      & $\ge 22.7$ &   $\le 29.9$   & $\le$ $-$7.2 	&  33.56(5)  &  $-$3.52(5)  \\ % &  	\\
 PSR J1357$-$6429  &   3.8    &  36.49	& 2.5  & 1.84  &  21.82(25) &  29.40(11) & $-$7.09(11)	&31.7(1)  &  $-$4.8(1)  \\ % & (5)\\  
     &       &  	&   &   &  $\ge$22.7 &  $\le$29.05 & $\le -$7.44	&     \\% & (5)\\  
 Vela   &	  4.1 &  36.84	&0.293$^{+0.019}_{-0.017}$   & 0.170(16)	&  21.76(6) &  27.52(8) &	$-$9.32(8) & 31.48(15)  &  $-$5.36(15)  \\ %&(4)	\\
 PSR B0656+14 &   5.0    &  34.58	&0.288$^{+0.033}_{-0.027}$     & 0.09(6)  &  22.11(13)$^*$  &	 27.38(16) &	$-$7.2(16)& 30.25(1)  &  $-$4.33(1)  \\ % &(5)	\\
 Geminga  & 5.5 & 34.51 & 0.250$^{+0.120}_{-0.060}$ & 0.12(9) & 23.4(2)$^*$ & 26.7(3)& $-$7.8(3) & 30.32(4) & $-$4.19(4) \\ %& (5)\\
  
 \hline   
 \end{tabular} \\
 \label{t:IR}
%\tablenotetext{$^*$}{$K_s$ magnitude is based on the extrapolation of the SED created using the HST F110W and F160W band data.}\\
%\tablenotetext{$^{**}$}{Logarithm of nonthermal pulsar luminosity in the 0.5-8 keV band.}\\ 
\begin{tabular}{l}
$^*$ $K_s$ magnitude is based on the extrapolation of the SED created using the HST F110W and F160W band data. \\
$^{**}$ Logarithm of nonthermal pulsar luminosity in the 0.5--8 keV band.\\
\end{tabular}
 %observations; \\
 %(1) \cite{2003A&A...406..639S}; (2) \cite{2012A&A...544A.100M}; (3) \cite{2006ApJ...644.1056K}; (4) \cite{2003A&A...406..645S}; (5)   \cite{2001A&A...370.1004K}
\end{table*}

The latter statement is also supported  by the   analysis of the  pulsar spindown power 
transformation efficiencies to the X-ray  and $K_s$-band emission.  They are described by the parameters $\eta_{X} \equiv L_{\rm X}/\dot{E}$ and  
$\eta_{K_s} \equiv L_{K_s}/\dot{E}$, where $L_{\rm X}$    and $L_{K_s}$ are the X-ray nonthermal  and $K_s$  luminosities, respectively.
In Table~\ref{t:IR}, we collected the related data for eight pulsars, including J1357, observed in the near-IR and X-rays. 
Only five of them have firm near-IR counterparts,  while  a near-IR counterpart candidate for PSR B1509$-$58   
\citep{Kaplan06-B1509-nIR} has not yet been confirmed and only an upper limit of the near-IR flux   was placed 
for PSR J1124$-$5916  \citep{zhar13}.     
In Fig.~\ref{fig:eff-eff}, we plot  Log $\eta_{K_s}$ vs.  Log $\eta_{X}$.
The plot demonstrates  a strong correlation between  these parameters. 
The correlation coefficient is 0.99. The positions 
of four firm IR pulsars in the Log $\eta_{X}$--Log $\eta_{K_s}$ plane can be fitted by a linear regression. 
The best fit and its 1$\sigma$ uncertainty are shown  by the dashed line and the grey filled region, respectively. 
The regression parameters  are also presented in the plot. 
The only pulsar, which shows a clear (about 3$\sigma$) excess of the near-IR efficiency over the fit, 
is PSR B0656$+$14. Note, that in the linear scale it overshoots the best fit by an order of magnitude.     
\citet{0656Durant} discussed that the excess could be due to the presence of a passive post-supernova 
fall-back disk around the pulsar. 
 However, one cannot exclude that the peculiar near-IR efficiency 
is an intrinsic  property of the PSR B0656$+$14 magnetosphere emission.     

It is also seen from Fig.~\ref{fig:eff-eff} that if the object $C'$ is confirmed as the J1357 near-IR counterpart, 
it will be another example of a pulsar with such a peculiar near-IR efficiency. 
This can be important for understanding the origin 
of the near-IR and optical emission of pulsars. Alternatively, the position  of the detection limit of our observations 
(Fig.~\ref{fig:eff-eff})  
with respect to the best fit  implies   that  J1357 may be  about two magnitudes 
fainter than $C'$.

%% Av=1.86; Ar=1.52, Ai=1.2

Unfortunately, the dereddened optical flux upper limits of the candidate obtained from the VLT data, 
$F_V\la2.9$ $\mu$Jy, $F_R\la2.6$ $\mu$Jy and $F_I\la 3.8$ $\mu$Jy,  overshoot 
the X-ray extrapolation and thus are uninformative.   
Observations of the J1357 field in the optical  using high spatial resolution imaging instruments, like the \textit{HST},
would be useful to verify the pulsar nature of the object $C'$  by extending its SED to shorter wavelengths. 
Further proper motion measurements
in the radio and near-IR would be also valuable.

\section*{Acknowledgements}
We thank the anonymous referee for useful comments. The work of DZ, AD and AK was supported by RFBR according to the research project No. 14-02-31600 mol\_a.
The work of YS was partially supported by RFBR, research projects No. 13-02-12017 ofi\_m and 
14-02-00868\_a. SZ acknowledges support from CONACYT 151858  project.
REM acknowledges support by  the BASAL Centro de Astrof\'isica y Tecnologias Afines (CATA) PFB--06/2007.
Image Reduction and Analysis Facility ({\tt IRAF}) is distributed by the National Optical Astronomy Observatories,
which are operated by the Association of Universities for Research
in Astronomy, Inc., under cooperative agreement with the National Science Foundation.
This research has benefited from the M, L, T and Y dwarf compendium housed at DwarfArchives.org.
%%%%%%%%%%%%%%%%%%%%%%%%%%%%%%%%%%%%%%%%%%%%%%%%%%



\bibliographystyle{mnras}
\bibliography{1357NACO_arxiv} % if your bibtex file is called example.bib


%%%%%%%%%%%%%%%%%%%%%%%%%%%%%%%%%%%%%%%%%%%%%%%%%%

%%%%%%%%%%%%%%%%% APPENDICES %%%%%%%%%%%%%%%%%%%%%

%%%%%%%%%%%%%%%%%%%%%%%%%%%%%%%%%%%%%%%%%%%%%%%%%%

% Don't change these lines
\bsp	% typesetting comment
\label{lastpage}
\end{document}